\documentclass{elsart}

\usepackage{graphicx}
\begin{document}

\begin{frontmatter}
\title{Crystal vs Glass Formation in Lattice Models with Many Coexisting Ordered Phases}

\author[USP]{M\'ario J. de Oliveira},
\author[CNR,INFM]{Alberto Petri},
\author[USP]{T\^ania Tom\'e}

\address[USP]
{Instituto de F\'{\i}sica, Universidade de S\~ao Paulo, Caixa Postal 66318,
05315-970 S\~ao Paulo, S\~ao Paulo, Brazil}

\address[CNR]{CNR, Istituto "O.M. Corbino",
via del Fosso del Cavaliere 100, 00133 Roma, Italy }
\address[INFM]{Unit\`a INFM, Universit\`a  La Sapienza, p.le Aldo Moro
2, 00185 Roma, Italy}

\begin{abstract}
We present here new evidence that
after a quench  the planar Potts model on the square lattice relaxes towards a glassy state if
the number of states $q$ is larger than four. By extrapolating the finite size data we compute 
the average energy of these states for the infinite system with  periodic 
boundary conditions, and find that it is comparable with that previously found 
using fixed boundary conditions.
We also report preliminary results on the behaviour of these states in the presence of 
thermal fluctuations.
\end{abstract}

\end{frontmatter}

\section{Introduction}

A variety of lattice models has been devised in the last years in order to
account for the freezing of liquids into a disordered state rather than
ordering into their crystalline forms after a sudden cooling below the solidification
temperature. Most of these models are based on a certain amount of quenched
disorder which prevents them from ordering \cite{crisanti03}. Others are provided of
dynamical \cite{ritort03} constrains aimed to the same purpose.
These models have been proven very useful for discovering and understanding
many important  features of the disordered systems. However they cannot reveal
the mechanisms that prevent crystallisation in systems with ordered ground
states.

Very recently departure from crystallisation has been observed even in some
models which do not possess any of the above ingredients and that on the contrary have
crystalline ground states \cite{petri03}. Strong metastability,
typical of the transition from liquid to glass, has 
been first observed in a model of Ising spins on a cubic
lattice  \cite{lipowski97,lipowski00} where the product of the four spins of each
plaquette contributes to the total energy with a same coupling constant
($4$-spin, or FSIM  model).
The model, which at equilibrium shows a first order transition, also displays
stretched exponential relaxation and
aging properties at low temperatures \cite{swift00}.

An interesting question concerns the origin of this behaviour and its
relationship with models that relax into ordered ground states. In some
recent works \cite{deoliveira02a,deoliveira03a,deoliveira03b}  we have
investigated the quench of two kinds of simple models on the square lattice, 
i.e.
exclusion models and Potts model, and found that under appropriate
circumstances also these models  may relax to glassy rather than to crystalline
or polycrystalline phases. The glassy phase is distinct from the
polycrystalline phase in that the length of the line separating different
domains is of the order of the number of sites in the system, thus being
different from the usual coexistence of different phases at thermodynamic
equilibrium. The behaviour is determined by the number $q$ of (equivalent)
ground states that the system possesses: in both kinds of models a glassy phase is
attained for $q >q_c$, whereas polycrystalline states are attained for $2 < q <
q_c$. A  remarkable point is that this change from ordering to glassiness in
the relaxation process, corresponds to the change from second to first order
transition in the corresponding equilibrium diagram, which occurs
at the same $q_c$ for both kind of systems.

\begin{figure}
\begin{center}
\includegraphics*[scale=0.20,angle=-90]{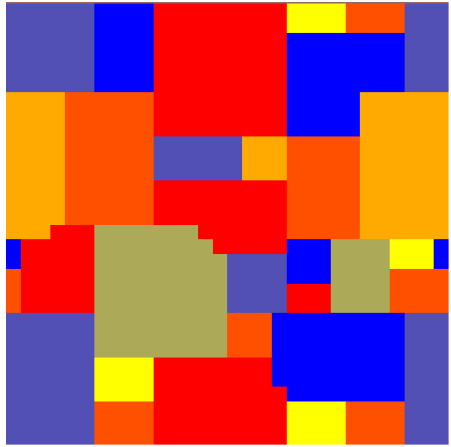}
\caption{Blocked configuration in a $7$-state Potts model quenched at $T=0$. The
{\sf T} shaped domain junctions are energetically stable.}
\end{center}
\end{figure}

The onset of a glassy phase  is not in contradiction with thermodynamics, according to which the
system evolves towards a uniform phases. In fact it is obtained under the
prescription of taking the thermodynamic ($L \rightarrow \infty$) limit before
the ergodic ($t \rightarrow \infty$) limit.  Thermodynamics
would require the ergodic limit to be taken before the thermodynamic limit.

Here we present some new findings. The next section reports
some results on the Potts model with different $q$ when subject to a quench at
$T=0$ with periodic boundary conditions (PBC). Section three deals with the
onset and the stability of the glassy phase at finite temperature, and  all  
results are summmarised in section four.

\section{Periodic boundary conditions  at $T=0$}

The Potts model with $q$ states may be defined by the Hamiltonian:
\begin{equation}
H=\sum_{ij} (1-\delta_{s_i,s_j}),
\end{equation}
where the sum is on nearest neighbour pairs, $s_i$ is the
state variable of the site $i$ which can assume the values $1,2,\dots,q$,
and $\delta_{a,b}$ is the Kronecker's function.

It is known since a long time that after a quench 
from high temperature to $T=0$
the system  relaxation  slows down and eventually stops when single spin 
dynamics is employed. 
This is due to the formation of locally
stable configurations, like the one shown in Fig.1, which in the absence of thermal fluctuations, keep the system's 
dynamics pinned. 
To overcome the
problem  of finite size effects, in our previous work 
\cite{deoliveira02a,deoliveira03a,deoliveira03b}
the quench was performed by imposing fixed boundary conditions (FBC), thus
forcing the system to finally fall into one of its ordered ground states. The
glassy phase was then predicted by the increase of the relaxation time with the
system size and the length of the line separating the different domains was 
obtained by a suitable extrapolation to $L=\infty$ and $t=\infty$.
\begin{center}
\begin{tabular}{|c|c|c|}
  \cline{1-3}
  q & PBC & FBC \\ \hline
$  3$ &$ 0.00$ &$ 0.000$ \\
$  5$ &$ 0.07$ &$ 0.060$  \\
$  7$ &$ 0.11$ &$ 0.105$  \\\hline
\end{tabular}

Table 1
\end{center}
 Blocked configurations observed with PBC have energies that, besides to
 depend on the system size, 
fluctuate strongly from one realization to the other.
Maybe for these reasons they have been generally considered some kind
of finite size artifact and, to our knowledge, they have never been investigated in a
systematic way. We have performed several quenches of the Potts model on square
lattices with increasing size $L$, using PBC and different values of $q$. For
each $L$, the average energy per site $u$ of the blocked configurations obtained 
is
reported in Fig.2, together with its statistical fluctuation, for $q=3,5$ and $7$.
While it shows to approach an asymptotic value for $q=3$, this does not seem
the case for $q=5$ and $7$.

\begin{figure}
\begin{center}
\includegraphics*[scale=0.40,angle=-90]{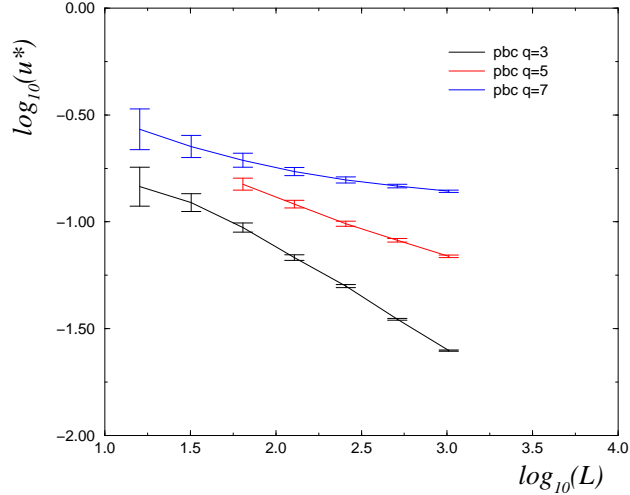}
\caption{Average energy reached by the system after a quench from $T=\infty$
to $T=0$
as a function of the system linear size $L$.
Curves for $q=5$ and $q=7$ display a tendency to saturate, not shown by the curve
for $q=3$.}
\end{center}
\end{figure}

In order to attempt quantitative predictions the same quantity has been plotted
on a different scale in Fig.3. The horizontal scale is chosen in such a way to
make the lines as straight as possible. This happens by choosing $1/\sqrt L$ as
plotting variable, but we have no theoretical argument for this choice.
Extrapolation to $L=\infty$ yields the average energy $u^*$ expected for the
blocked configurations in the thermodynamic limit. A comparison of these values with
those obtained by using FBC \cite{deoliveira02a,deoliveira03a,deoliveira03b}
is reported in Table $1$ and confirms our previous conclusions, that is
for $q > 4$ the quench leads to a glassy phase: the length of the
line separating different domains contribute with an energy which does not
vanish in the limit of infinite system. We did not investigate the case $q=4$
which is critical  \cite{deoliveira02a,deoliveira03a,deoliveira03b} and might
require logarithmic corrections.

\begin{figure}
\begin{center}
\includegraphics*[scale=0.40,angle=-90]{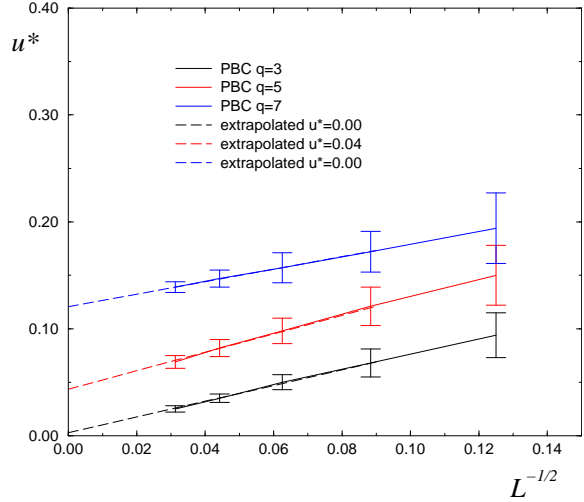}
\caption{The limit energy per site $u*$ displayed in Fig.2 appears to behave linearly when
plotted as function of $1/\sqrt(L)$. Extrapolation to $L=\infty$ yields values
compatible with those obtained when using FBC, reported in Table I.  }
\end{center}
\end{figure}

\section{Stability against thermal fluctuations}

Our previous work  \cite{deoliveira02a,deoliveira03a,deoliveira03b} concerned
the behaviour of Potts and exclusion model after a quench at $T=0$
\cite{atermic}. Here we report some preliminary results on two different cases. The first
is the case in which the quench is performed at a low but finite temperature.
The energy decay of the Potts model with $q=7$ after a quench at $T=0.1$, is
plotted in Fig.4 for different system size as a function of the expected domain
size. As this latter increases, two main trends may be singled out: $i)$ the
relaxation to the ground state (forced by the use of FBC) takes places at later
times; $ii)$ it starts at increasing values of the energy. Although yet preliminary,
these evidences indicate that the system would relax to a state with
finite energy even when quenched at a temperature that is about $0.13$
of the critical temperature. More work in this direction is still needed for 
drawing quantitative conclusions.

\begin{figure}
\begin{center}
\includegraphics*[scale=0.40,angle=-90]{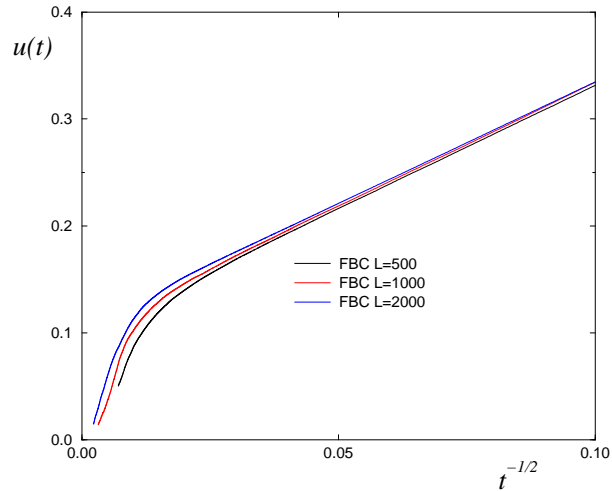}
\caption{The relaxation to equilibrium after a quench from $T=\infty$ to
$T=0.1$. Relaxation to the ground state takes place later and the energy decreases
less, when the system size increases.}
\end{center}
\end{figure}

The second case concerns the investigation of the time taken by  the system to
escape from a metastable configuration because of thermal fluctuations. 
The system is initially at $T=\infty$; then it is quenched to $T=0$ and left to
relax into a metastable state. This is numerically checked by  verifying that 
the system energy cannot decrease further by changing state to a single site.
Then temperature is raised again to a finite values $T_e$, and the time needed to overcome 
the energy barrier 
is computed as the time needed to reach the thermal equilibrium
energy corresponding to  that temperature. 

Figure 5 reports the time distribution for $q=7$ and $T_e=0.4$, which is more
than half the critical temperature.  Histograms are computed over $10,000$ realizations
for $L=16$ and $32$, and on more than $1000$ realizations for $L=64$.
Owing to the large statistics required, only results for small systems are available
at present, more work being in progress. The data show that the distribution of 
the logarithm of escape times
is approximately Gaussian, but a distortion towards large times seems to develop 
for $L=64$. It should  be noted that the number of quenches leading to states with the energy
of the ground state vanishes and that the typical escape time, shown in the inset, grows about exponentially 
when the system size increases. It is expected to be proportional to the 
height of the free energy barriers separating 
metastable phases by stable ones. In the Potts model this height
grows as $\exp(L^{d-1})$ \cite{lee91}.

\begin{figure}
\begin{center}
\includegraphics*[scale=0.40,angle=-90]{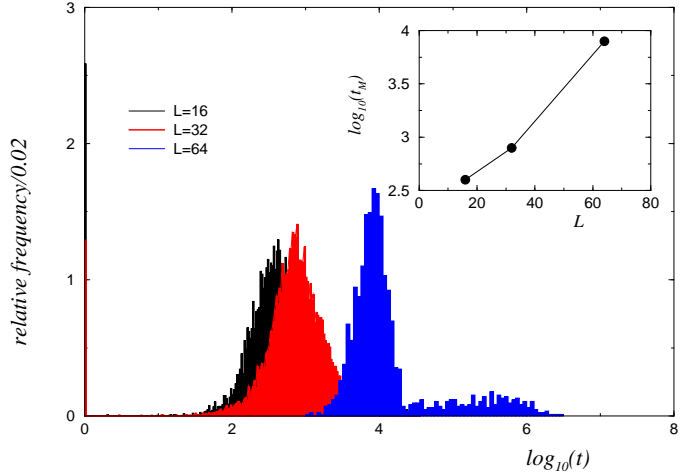}
\caption{Distribution of the logarithm of escape times from a metastable state, for the 
Potts model with $q=7$, when the temperature is raised from $T=0$ to $T=0.4$. 
In the inset it is plotted the typical time.  Note the log scale on the time axis.}
\end{center}
\end{figure}
\section{Summary}
We have reported here some recent results on the onset of glassy states  
after the quench of the Potts model on a square lattice. 
By a method different from previous work,
they confirm that these states occur when the number of ground states $q$ is larger 
than four 
\cite{deoliveira02a,deoliveira03a,deoliveira03b}.
By extrapolating the computed average energies  
we found values for the infinite system that are compatible
with those found  with the previous method for the three values of $q$ 
investigated.  

Preliminary results show that the glassy state is attained also by quenching at small but finite 
$T$, and that for $q=7$ the time needed by the system to exit from a local energy minimum by
thermal fluctuations seems to increase  exponentially with the system size
for temperatures of the order of half the critical temperature.


\begin{thebibliography}{99}
\bibitem{atermic} Actually, temperature does not appear in exclusion
models in explicit form, but only in combination with the chemical potential
$\mu$, which determines the deposition and evaporation rates of particles,
through the ratio $\mu/T$. Thus the limit $T=0$ is indistinguishable from
$\mu\rightarrow \infty$ and for this reason these systems are sometimes called
athermal.
\end{thebibliography}
\end{document}